\documentclass[RNAAS]{aastex62}

\submitjournal{RNAAS}

\shorttitle{On the scatter of dwarfs in present-day MZRs}
\shortauthors{Engler, Lisker \& Pillepich}

\begin{document}

\title{On the scatter of the present-day stellar metallicity-mass relation\\ of cluster dwarf galaxies}

\author{Christoph Engler}
\author{Thorsten Lisker}
\affiliation{Astronomisches Rechen-Institut, Zentrum f\"ur Astronomie der Universit\"at Heidelberg, \\M\"onchhofstra\ss e 12-14, D-69120 Heidelberg, Germany}

\author{Annalisa Pillepich}
\affiliation{Max-Planck-Institut f\"ur Astronomie, K\"onigstuhl 17, 69117 Heidelberg, Germany}
\affiliation{Harvard-Smithsonian Center for Astrophysics, 60 Garden Street, Cambridge, MA 02138}


\section{Introduction}
The mass-metallicity relation (MZR) is one of the fundamental relations of galaxies, manifesting itself observationally for cluster galaxies in color-magnitude relations \citep{smithcastelli2008} or relations of absorption line indices with stellar mass \citep{toloba2014}. Satellite galaxies, however, have been found to be more metal-rich than centrals of the same mass, both in nature \citep{pasquali2010} as well as in cosmological simulations \citep{bahe2016, genel2016}. In this research note, we examine the scatter of the relation between stellar \text{mass $M_*$} and stellar \text{metallicity $Z_*$} for cluster dwarf galaxies in the cosmological simulation Illustris \citep{vogelsberger2014a, vogelsberger2014b, genel2014, nelson2015}. We focus on galaxies in the stellar mass range $10^8 \text{ M}_\odot \leq M_* < 10^{10} \text{ M}_\odot$ in a cluster of total mass $3.8 \cdot 10^{14} \text{ M}_\odot$ at redshift $z=0$. This mass range is particularly interesting, since early-type galaxies were found to have large ranges of enrichment \citep{michielson2008}, stellar ages \citep{paudel2010a} and rotational support \citep{rys2014}. We find the MZR to exhibit the smallest intrinsic scatter at times of peak stellar mass, i.e. when the galaxies were at their respective maximum stellar mass. This suggests stellar mass stripping to be the primary effect for the rather broad relation observed at present.\\

\section{Results}
\begin{figure}[tbp]
\centering
\includegraphics[width=.95\textwidth]{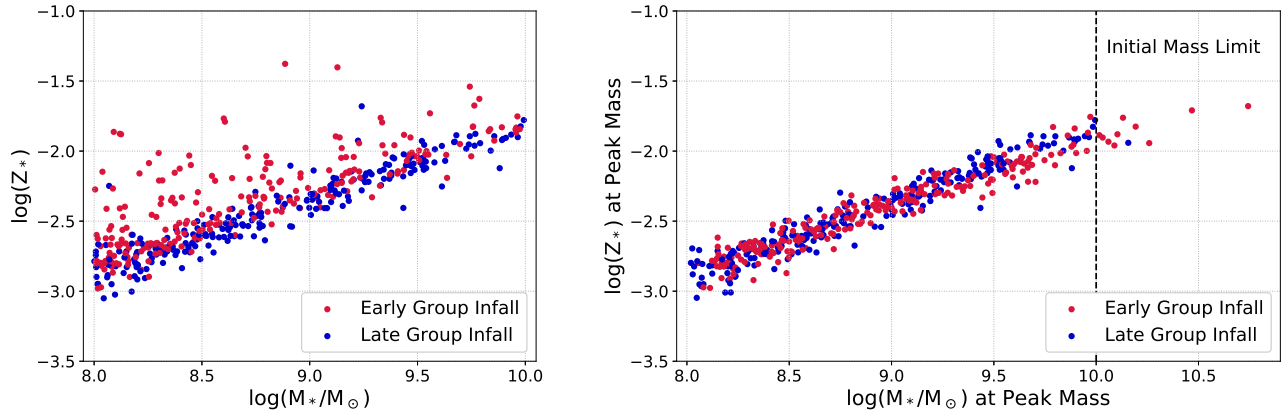}
\caption{MZRs for Illustris low-mass galaxies residing in a $z=0$ cluster of $3.8 \cdot 10^{14} \text{ M}_\odot$, color-coded by early ($z \geq 0.76$, $\text{lookback time} \geq 6.65 \text{ Gyr}$) and late infall into a group of at least $10^{12} \text{ M}_\odot$. Left: MZR of all galaxies at $z=0$. Right: MZR using mass and metallicities at the galaxies' individual times of peak stellar mass.}
\label{fig:mzr}
\end{figure}

The left panel of \text{Figure \ref{fig:mzr}} shows the MZR for the low-mass galaxies of the cluster, which experienced a major merger $3.5 \text{ Gyr}$ prior at $z=0.31$. The relation depicts a broad scatter with a pronounced tail towards its high-metallicity side ($0.180 \text{ dex}$ at $1\sigma$). Most high-metallicity outliers fell early into a group environment, which we define here as halos of total mass $\geq 10^{12} \text{ M}_\odot$.
The subsample of galaxies with early group infall yields a scatter of $0.210 \text{ dex}$, while the scatter for late infallers is at $0.099 \text{ dex}$. Clearly, the environment and the time the galaxies spent therein is connected to the broadening of the MZR.

In the right panel of \text{Figure \ref{fig:mzr}} we consider the galaxies at their \textit{individual} times of peak stellar mass. With a scatter of $0.077 \text{ dex}$, this relation is narrower than any of the equal-time relations we examined, such as the MZR of the galaxies' progenitors at $z=1$ (scatter: $0.106 \text{ dex}$) or the time at which $50\%$ of the progenitors had experienced group infall ($z=0.76$, scatter: $0.109 \text{ dex}$). Observationally derived MZRs necessarily exhibit larger scatter, since observers only have access to equal-time MZRs, cannot distinguish the galaxies by their respective infall times, and cannot easily reconstruct the galaxy evolution.

We examine the time evolutions of stellar masses and stellar metallicities and find that stripping of stellar mass (more pronounced in more massive cluster hosts) is the primary process responsible for the broad MZR at present-day times. However, for about $40\%$ of galaxies in the high-metallicity tail, mass stripping coincides with an increased enrichment of stellar metallicity. This is possibly caused by the stripping of low-metallicity stars in the galaxy outskirts \citep{bahe2016,genel2016} or perhaps gas inflows inducing starbursts \citep{janz2016}.\\

\newpage
\section{Discussion}
Our findings rely on the realism of Illustris galaxies. However, galaxy sizes in Illustris are generally too large by a factor of \text{\raise.17ex\hbox{$\scriptstyle\sim$}}2-3 for galaxies with  $M_* \la 10^{10.7} \text{ M}_\odot$ \citep{nelson2015, snyder2015}, compared to simulations of isolated dwarf galaxies \citep{vandenbroucke2016} or observed dwarfs in the Virgo cluster \citep{janz2014}. Having their stars spread over such an extended radius makes the low-mass galaxies even more susceptible to external influences \citep{bialas2015}, thereby increasing the efficiency of stellar mass stripping.

It stands to question how strongly the MZR has been affected by these processes and how different it might appear with galaxies in the successor simulation IllustrisTNG \citep{marinacci2017, naiman2018, nelson2018, pillepich2018b, springel2017}, which incorporates improved galaxy models resulting in smaller and more realistic galaxy sizes at low masses \citep{weinberger2017, pillepich2018a}. First studies show signs of systematic chemical pre-processing in the gas-phase metallicities of infalling galaxies \citep{gupta2018}. Comparing the effects on the MZR in Illustris and IllustrisTNG as well as the further importance of mass stripping and metallicity enrichment for the evolution of low-mass galaxies remain to be investigated more comprehensively.\\


\end{document}